\documentclass[letterpaper,english]{article}
\usepackage[T1]{fontenc}
\usepackage[latin9]{inputenc}
\usepackage{amsmath}
\usepackage{amssymb}
\usepackage{graphicx}

\makeatletter

\usepackage{spconf}

\usepackage{enumitem}
\usepackage{amsthm}
\usepackage{colortbl}
\usepackage{array}
\newcolumntype{C}[1]{>{\centering\arraybackslash}p{#1}}

\def\hlinewd#1{%
\noalign{\ifnum0=`}\fi\hrule \@height #1 %
\futurelet\reserved@a\@xhline}
\newcolumntype{"}{@{\hskip\tabcolsep\vrule width 1pt\hskip\tabcolsep}}

\title{LOW-LATENCY LIST DECODING OF POLAR CODES WITH DOUBLE THRESHOLDING}
\name{YouZhe Fan$^{\star}$, Ji Chen$^{\star}$, ChenYang Xia$^{\star}$, Chi-ying Tsui$^{\star}$, Jie Jin$^{\dagger}$, Hui Shen$^{\dagger}$, and Bin Li$^{\dagger}$}

\address{$^{\star}$ Department of Electronic and Computer Engineering, the HKUST, Hong Kong\\
    $^{\dagger}$Communications Technology Research Lab., Huawei Technologies, P. R. China}

\@ifundefined{showcaptionsetup}{}{%
 \PassOptionsToPackage{caption=false}{subfig}}
\usepackage{subfig}
\makeatother

\usepackage{babel}
\begin{document}
\ninept
\maketitle
\begin{abstract}
For polar codes with short-to-medium code length, list successive
cancellation decoding is used to achieve a good error-correcting performance.
However, list pruning in the current list decoding is based on the
sorting strategy and its timing complexity is high. This results in
a long decoding latency for large list size. In this work, aiming
at a low-latency list decoding implementation, a double thresholding
algorithm is proposed for a fast list pruning. As a result, with a
negligible performance degradation, the list pruning delay is greatly
reduced. Based on the double thresholding, a low-latency list decoding
architecture is proposed and implemented using a UMC 90nm CMOS technology.
Synthesis results show that, even for a large list size of 16, the
proposed low-latency architecture achieves a decoding throughput of
220 Mbps at a frequency of 641 MHz.
\end{abstract}

\section{Introduction}

\label{sec:intro}

Successive cancellation decoding (SCD) is proposed in \cite{Arikan}
for decoding polar codes, and its hardware implementation is extensively
studied in \cite{SSC}-\cite{ASSCC}. However, for polar codes with
short-to-medium code length, the error-correcting performance of the
SCD is unsatisfactory. To improve the performance, SCDs with multiple
codeword candidates are proposed. They are the list decoding \cite{list},
\cite{list_BUPT} and its variants \cite{stack_BUPT}-\cite{SD_BUPT}.
For a better performance, cyclic redundancy check (CRC) code is serially
concatenated with polar codes and the CRC bits are used to choose
the valid codeword from the list candidates \cite{list}, \cite{CRC_BUPT},
\cite{CRC_Bin}. As a result, the list decoding of polar codes achieves
or even exceeds the performance of Turbo codes \cite{ICC_BUPT} and
LDPC codes \cite{list}. However, this performance improvement is
at the cost of a larger list size (e.g., 16 or 32) and the increased
complexity highly desires an efficient list decoding architecture.
In this work, the efficient and low-latency implementation of the
list decoding is explored, aiming at promoting polar codes as a competitive
coding candidate in both error-correcting and implementation aspects.

The first list decoding architecture for polar codes is proposed in
\cite{TCASII_EPFL}. In \cite{ISCAS_Xiaohu}, the pre-computation
look-ahead technique \cite{look_ahead_parhi} is used in the list
decoding for a lower latency, while its memory size is tripled. In
\cite{TCASII_EPFL} and \cite{ISCAS_Xiaohu}, a small list size of
4 and 2 are used, respectively. When the list decoding decodes an
information bit, the number of the codeword candidates are doubled.
To maintain a reasonable decoding complexity, once the candidate size
exceeds the specific list size $\mathcal{L}$, some of the codeword
candidates have to be pruned. The common pruning strategy is to sort
the codeword candidates based on their metrics and keep the $\mathcal{L}$
best of them. However, the sorting operation incurs a large hardware
and timing complexity, especially when $\mathcal{L}$ is large. In
\cite{ISCAS_Lehigh}, a list decoding architecture with list size
of 8 is proposed, and a Bitonic sorting network is customized for
efficient sorting. Nevertheless, up to three pipeline stages are used
by the sorting architecture. As a result, to implement the list decoding
with large list size in hardware, list pruning architecture is critical,
especially to achieve a low decoding latency.

In this work, the list pruning architecture is optimized in both algorithmic
and architectural levels. Recently, instead using log-likelihood (LL)
to capture the metric of the list candidates, the LL ratio (LLR) representation
is used for the list decoding \cite{ICASSP_EPFL}-\cite{Asilomar_Parhi}.
Benefiting from the numerical accuracy and stability of the LLR, a
small and regular architecture of the memory and processing element
(PE) can be used for the list decoding \cite{ICASSP_EPFL}. Therefore,
in this work, LLR is used in the design of the low-latency pruning
architecture of the list decoder. Very recently in \cite{arxiv_EPFL}-\cite{TVLSI_Parhi},
borrowing some advanced techniques used in the SCD implementation
\cite{SSC}, the special constituent codes of polar codes are utilized
to reduce the latency of the list decoding. However, conventional
sorting strategies are still used for their list pruning and this
limits the latency reduction.

\section{RELATION TO PRIOR WORK}

\label{sec:format}

The main contributions of this work are outlined as follows:

\begin{enumerate} [leftmargin=*,topsep=0pt,itemsep=-1ex,partopsep=1ex,parsep=1ex]
\item Different from the previous works on list decoding \cite{list}-\cite{TVLSI_Parhi}, a double thresholding strategy (DTS) is proposed to replace the sorting strategy for list pruning.
\item In the architectural level, the architectures for DTS and threshold value update are proposed. As a result, even for a large list size, the logic delay of list pruning is very small.
\item A low-latency list decoding architecture for a large list size, i.e. 16, is implemented in the UMC 90nm CMOS technology. Its decoding latency is even smaller than that of list size of 8 \cite{ISCAS_Lehigh}.
\end{enumerate}

\section{LIST DECODING OF POLAR CODES}

\label{sec:pagestyle}

A length $N=2^{n}$ polar code with rate $R=K/N$ is specified by
the generator matrix $\mathbf{G}_{N}$ and a frozen set $\mathcal{A}^{c}\subset\left\{ 0,1,\ldots,N-1\right\} $
of cardinality $\left|\mathcal{A}^{c}\right|=N-K$. A source word
of polar codes is denoted as $\mathbf{u}_{N}$, and $\mathbf{u}_{N}\in\left\{ 0,1\right\} ^{N}$.
It consists of $K$ information bits $u_{i}$ ($i\notin\mathcal{A}^{c}$)
and $N-K$ frozen bits $u_{i}$ ($i\in\mathcal{A}^{c}$). The information
bit is used to deliver the data, while the frozen bit is set to a
value, e.g., 0, pre-known by the decoder. If the $r$-bit CRC is used,
the last $r$ information bits take the CRC of the previous $K-r$
bits. In the encoder, the codeword $\mathbf{x}_{N}\in\left\{ 0,1\right\} ^{N}$
is generated as $\mathbf{x}_{N}^{T}=\mathbf{u}_{N}^{T}\mathbf{G}_{N}$
and sent over the physical channels.

Let $\mathbf{y}$ be the noise corrupted signal of $\mathbf{x}_{N}$
at the receiver. The LLRs input to the decoder are given as
\begin{equation}
L_{i}^{0}=\log\left[\Pr\left(\mathbf{y}|x_{i}=0\right)\right]-\log\left[\Pr\left(\mathbf{y}|x_{i}=1\right)\right]
\end{equation}
for $i=0,1,\ldots,N-1$. The decoding process of polar codes can be
illustrated by two trees: the decoding tree and the scheduling tree.
Fig. 1 shows an example of the trees for $N=4$. The decoding tree
of a length-$N$ polar code is a depth-$N$ binary tree, with $u_{i}$
mapped to the nodes at depth $i+1$. Its root node represents a null
state. A path from the root node to a depth-$i$ node represents a
subvector $\left[u_{0},u_{1},\ldots,u_{i-1}\right]$ of the source
word $\mathbf{u}_{N}$, and is named as the decoding path $p^{i}$.
Specifically, a path from the root node to the leaf node of the decoding
tree represents a source word $\mathbf{u}_{N}$ of polar codes, and
the value of each bit of $\mathbf{u}_{N}$ is shown in the corresponding
node lying at this decoding path. Notice that, if $u_{i}$ is a frozen
bit, it only assumes 0. Hence, the right sub-tree rooted at the depth-$\left(i+1\right)$
node can be pruned, as the source words included in it are not valid.
For example, if $\mathcal{A}^{c}=\left\{ 0\right\} $, the gray sub-tree
in Fig. 1(a) is pruned.
\begin{figure}
\subfloat[Decoding tree and decoding path]{\includegraphics{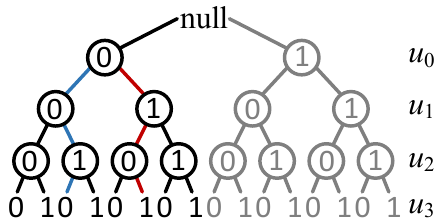}

}\subfloat[Scheduling tree for SCD]{\includegraphics{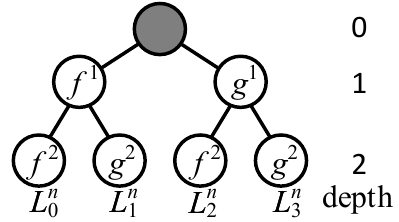}

}\vspace{-2mm}

\caption{List decoding example of polar codes with $N=4$ and $\mathcal{L}=2$}
\vspace{-5mm}
\end{figure}

Decoding the polar codes can be treated as a search problem in the
pruned decoding tree. The conventional SCD performs a depth-first
search. Given a partial decoding path $\left[u_{0},u_{1},\ldots,u_{i-1}\right]$,
the SCD generates the LLR of bit $u_{i}$, denoted as $L_{i}^{n}$.
If $i\in\mathcal{A}^{c}$, $u_{i}$ is decoded as $\hat{u}_{i}=0$
irrespective of $L_{i}^{n}$. Otherwise, a ML decision is made for
the information bit $u_{i}$ ($i\notin\mathcal{A}^{c}$), and is given
by
\begin{equation}
\hat{u}_{i}=\Theta\left(L_{i}^{n}\right)=\begin{cases}
\begin{array}{l}
0\\
1
\end{array} & \begin{array}{l}
L_{i}^{n}\geq0\\
L_{i}^{n}<0
\end{array}\end{cases}
\end{equation}
Based on the decision rule in (2), single decoding path from the root
to the leaf is obtained in the decoding tree, e.g., the red path in
Fig. 1(a), and it is the source word $\mathbf{\hat{u}}_{N}$ decoded
by the SCD.

For a better error-correcting performance, a breadth-first search
is performed by the list decoding. To constrain the searching complexity,
a list size $\mathcal{L}$ is set. Let $\mathcal{L}$ decoding paths
at depth $i$ of the decoding tree be denoted as $p_{l}^{i}=\left[u_{0}^{l},u_{1}^{l},\ldots,u_{i-1}^{l}\right]$,
$l=0,1,\ldots,\mathcal{L}-1$. For each path candidate $p_{l}^{i}$,
a path metric is associated with it and denoted as $pm_{l}^{i}$.
When decoding the information bit $u_{i}$, $\mathcal{L}$ decoding
paths are extended to $2\mathcal{L}$ paths. From \cite{ICASSP_EPFL},
the path metrics of the two extensions of the path $p_{l}^{i}$ are
given by
\begin{equation}
pm_{l}^{i+1}\left(u_{i}\right)=pm_{l}^{i}+\log\left[1+e^{\left(2u_{i}-1\right)L_{i}^{n}}\right]
\end{equation}
where $u_{i}$ assumes 0 and 1, corresponding to the left and right
extensions of $p_{l}^{i}$. In the hardware \cite{ICASSP_EPFL}, (3)
is approximated by
\begin{equation}
pm_{l}^{i+1}\left(u_{i}\right)=\begin{cases}
\begin{array}{l}
pm_{l}^{i}\\
pm_{l}^{i}+\left|L_{i}^{n}\right|
\end{array} & \begin{array}{l}
\textrm{if }u_{i}=\Theta\left(L_{i}^{n}\right)\\
\textrm{if }u_{i}\neq\Theta\left(L_{i}^{n}\right)
\end{array}\end{cases}
\end{equation}
The operation in (4) is denoted as path metric update (PMU). Based
on the $2\mathcal{L}$ $pm$s from PMU, $\mathcal{L}$ extended paths
with the smallest $pm$s are chosen and they are the paths at depth
$i+1$, i.e., $p_{l}^{i+1}$, $0\leq l<\mathcal{L}$. This operation
is named as the list pruning operation (LPO). 

From (3) or (4), it can be seen that the PMU needs the knowledge of
$L_{i}^{n}$ and it is generated by the SCD. The SCD operation can
be described by the scheduling tree shown in Fig. 1(b). The scheduling
tree of a length-$N$ polar code is a depth-$n$ balanced binary tree.
It consists of two kinds of nodes: $f$ node and $g$ node. The functions
included in one node can be evaluated in one clock cycle. Generally,
the depth-first traversal of the scheduling tree completes decoding
one codeword, and the $i^{\textrm{th}}$ leaf node outputs the LLR
$L_{i}^{n}$. In the list decoding, $\mathcal{L}$ SCDs are deployed
and executed in parallel. When they reach the leaf node, the SCDs
are stalled and the PMU calculates $2\mathcal{L}$ path metrics from
$\mathcal{L}$ $L_{i}^{n}$s with (4). After that, the LPO chooses
the best $\mathcal{L}$ decoding paths. The SCD cannot be restarted
until the LPO finishes, because the subsequent SCD operations need
the knowledge of the updated decoding paths $p_{l}^{i+1}$. Therefore,
the delay of the PMU and the LPO result in an increased latency of
the list decoding. 

From (4), the PMU is implemented with an adder array and its logic
delay is small. However, a sorting strategy is used for the LPO in
the conventional list decoding architecture \cite{TCASII_EPFL}-\cite{ICASSP_EPFL}.
For a shorter delay, a parallel sorting architecture \cite{TVLSI_EPFL}
is used in \cite{TCASII_EPFL} and \cite{ICASSP_EPFL}. However, its
hardware complexity is $\mathcal{O}\left(\mathcal{L}^{2}\right)$,
and hence it becomes inefficient for large $\mathcal{L}$. On the
other hand, a Bitonic sorting network is used in \cite{ISCAS_Lehigh},
and its delay also scales with $\mathcal{L}$. Next, to achieve a
short-delay LPO and hence a low-latency decoder, a double thresholding
algorithm and its corresponding architecture are proposed.

\section{LOW-LATENCY LIST DECODING IMPLEMENTATION}

\label{sec:typestyle}

\subsection{Double Thresholding Strategy}

\label{sssec:subsubhead-1}

\begin{figure}
\subfloat[$RT=pm_{\mathcal{L}-1}^{i}$]{\includegraphics{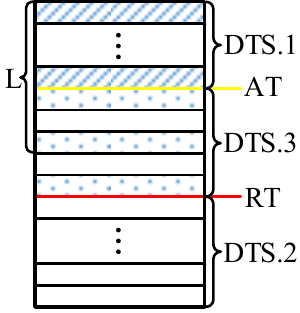}

}\subfloat[A tighter RT]{\includegraphics{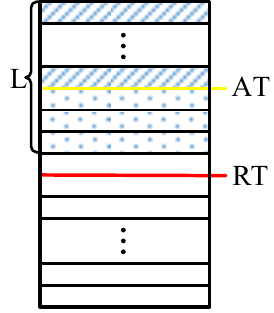}

}\subfloat[Too tight RT]{\includegraphics{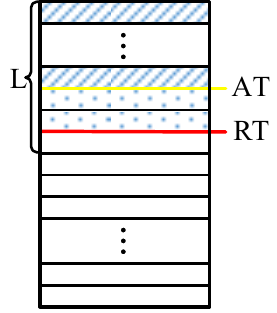}

}\vspace{-2mm}

\caption{Double thresholding strategy}
\vspace{-5mm}
\end{figure}
In this sub-section, a list pruning strategy with small logic delay
is introduced. Based on the $2\mathcal{L}$ path metrics from PMU,
it approximately finds the $\mathcal{L}$ smallest $pm$s and their
corresponding path extensions. To achieve it, the properties of the
$2\mathcal{L}$ path metrics in (4) are firstly studied and presented
in the following proposition. \newtheorem{prop}{Proposition}\begin{prop}Assume
$\mathcal{L}$ path metrics at depth $i$ of the decoding tree are
sorted and 
\begin{equation}
pm_{0}^{i}<pm_{1}^{i}<\cdots<pm_{l}^{i}<pm_{l+1}^{i}<\cdots<pm_{\mathcal{L}-1}^{i},
\end{equation}
and they are extended to $2\mathcal{L}$ path metrics with (4). If
the subset of $pm_{l}^{i+1}\left(u_{i}\right)$s smaller than $T$
is defined as 
\begin{equation}
\Omega\left(T\right)=\left\{ pm_{l}^{i+1}\left(u_{i}\right)|pm_{l}^{i+1}\left(u_{i}\right)<T\right\} ,
\end{equation}
then, the cardinality of $\Omega\left(T\right)$ for $T=pm_{l}^{i}$
satisfies
\begin{equation}
l\leq\left|\Omega\left(pm_{l}^{i}\right)\right|\leq2l.
\end{equation}
\end{prop}

Due to the space limitation, the proof of Proposition 1 is not shown.
Based on Proposition 1, the \textit{Double Thresholding Strategy}
(DTS) for list pruning is given as follows.\newtheorem*{dts*}{Double Thresholding Strategy}\begin{dts*}Assume
$\mathcal{L}$ path metrics at depth $i$ of the decoding tree follow
(5). To prune the $2\mathcal{L}$ path extensions at depth $i+1$,
two thresholds, i.e. Acceptance Threshold (AT) and Rejection Threshold
(RT), are defined and set as 
\begin{equation}
\left[AT,RT\right]=\left[pm_{\mathcal{L}/2}^{i},pm_{\mathcal{L}-1}^{i}\right].
\end{equation}
The path extensions at depth $i+1$ obey the following pruning rule:

\begin{enumerate} [leftmargin=*,topsep=0pt,itemsep=-1ex,partopsep=1ex,parsep=1ex]
\item if $pm_{l}^{i+1}\left(u_{i}\right)<AT$, the path extension is kept;
\item if $pm_{l}^{i+1}\left(u_{i}\right)>RT$, the path extension is pruned;
\item for path extensions with $AT\leq pm_{l}^{i+1}\left(u_{i}\right)\leq RT$, they are randomly chosen such that the list size remains to be $\mathcal{L}$.
\end{enumerate} 

\end{dts*}

Fig. 2(a) illustrates the DTS for LPO. Assume the $2\mathcal{L}$
extended $pm$s are sorted and the top path extension has the smallest
path metric. If the list is exactly pruned, the top $\mathcal{L}$
path extensions will be the decoding paths at depth $i+1$. However,
when the DTS is used, the shaded paths are reserved for depth $i+1$.
As shown in Fig. 2(a), from Proposition 1, DTS.1 ensures that at least
$\mathcal{L}/2$ best decoding paths are kept. Moreover, the number
of the reserved paths does not exceed $\mathcal{L}$. On the other
hand, since $\left|\Omega\left(pm_{\mathcal{L}-1}^{i}\right)\right|\geq\mathcal{L}-1$,
DTS.2 efficiently excludes the path extensions that are definitely
not in the set of the $\mathcal{L}$ best paths. Finally, when the
number of the paths kept by DTS.1 is smaller than $\mathcal{L}$,
DTS.3 will fill up the $\mathcal{L}$ path candidates. Notice that
the number of the pruned paths in DTS.2 is no greater than $\mathcal{L}$.
Therefore, DTS.3 is always used to fill up the decoding list.

From Fig. 2(a), the performance degradation of the DTS is due to DTS.3.
If RT is loose, some decoding path belongs to the $\mathcal{L}$ best
paths may not be chosen by DTS.3. To alleviate this, a tighter (smaller)
RT can be assumed. For example, the value of RT in (8) can be replaced
by $pm_{k}^{i}$ ($k<\mathcal{L}-1$). As shown in Fig. 2(b), by doing
so, the number of the candidates that DTS.3 can choose decreases,
and hence the probability that the chosen decoding path belongs to
the $\mathcal{L}$ best paths increases. However, from (7), when $RT=pm_{k}^{i}$
($k<\mathcal{L}-1$), it is possible that more than $\mathcal{L}$
decoding paths will be pruned by DTS.2. As a result, DTS.3 is not
always able to fill up the $\mathcal{L}$ path candidates, as depicted
in Fig. 2(c). Hence, if $RT$ value is too small, the performance
will become poor, as the decoding paths are aggressively pruned. Therefore,
an optimal value of $RT$ exits.
\begin{figure}
\begin{centering}
\subfloat[Architecture for AT]{\includegraphics{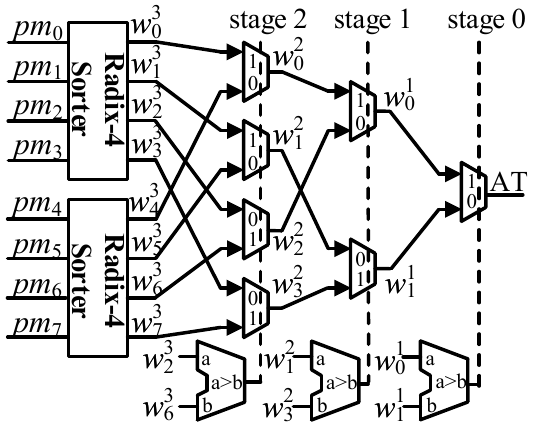}

}\subfloat[Architecture for RT]{\hspace{-1mm}\includegraphics{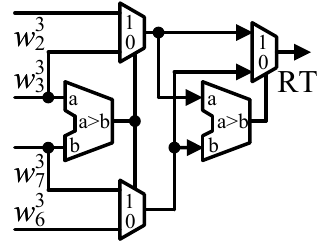}

}\vspace{-2mm}

\par\end{centering}

\caption{Threshold tracking architecture}
\vspace{-5mm}
\end{figure}

Finally, from the hardware implementation perspective, the complexity
of the DTS is much smaller than that of the conventional sorting strategy.
To implement DTS.1 and DTS.2, $4\mathcal{L}$ comparators are sufficient
and all the comparison operations can be executed in parallel as the
$pm$s are compared with the same fixed threshold values. To implement
DTS.3, the circuits based on the priority encoder are used. Most importantly,
due to the parallel nature of the DTS, the logic delay of the DTS
is much shorter than that of the full sorting strategy. As a result,
the PMU together with the DTS can be finished in one clock cycle.

\subsection{Threshold Tracking Architecture}

To support the DTS block, the values of AT and RT are needed. These
values are calculated by the \textit{Threshold Tracking Architecture}
(TTA) shown in Fig. 3. From Section 4.1, AT and RT used at depth $i+1$
of the decoding tree depend on the path metric at depth $i$. Therefore,
the TTA can be executed in parallel with the list decoding in extending
the path from depth $i$ to $i+1$. This leads to a relaxed timing
budget for TTA, and it can be executed in multiple cycles.

From (8), the TTA finds the median and the maximum values of the $\mathcal{L}$
input numbers. Finding the median is more complicated and its implementation
is based on the following property of the medians. \begin{prop}Assume
$W$ numbers $\left\{ w_{0},w_{1},\ldots,w_{W-1}\right\} $ satisfy
the following properties: 
\begin{equation}
\begin{cases}
\begin{array}{l}
w_{0}\leq w_{1}\leq\cdots\leq w_{m_{0}}\leq\cdots\leq w_{W/2-1}\\
w_{W/2}\leq w_{W/2+1}\leq\cdots\leq w_{m_{1}}\leq\cdots\leq w_{W-1}
\end{array}\end{cases}
\end{equation}
where $w_{m_{0}}$ and $w_{m_{1}}$ are the medians of $\left\{ w_{0},\ldots,w_{W/2-1}\right\} $
and $\left\{ w_{W/2},\ldots,w_{W-1}\right\} $, respectively. If the
median of $\{w_{0},w_{1},\ldots,$ $w_{W-1}\}$ is denoted as $w_{m}$,
then, 
\begin{equation}
\hspace{-1mm}\begin{cases}
\hspace{-2mm}\begin{array}{l}
w_{m}\in\left\{ w_{m_{0}},\ldots,w_{W/2-1},w_{W/2},\ldots,w_{m_{1}}\right\} \\
w_{m}\in\left\{ w_{m_{1}},\ldots,w_{W-1},w_{0},\ldots,w_{m_{0}}\right\} \\
w_{m}=w_{m_{0}}=w_{m_{1}}
\end{array} & \hspace{-6mm}\begin{array}{l}
w_{m_{0}}<w_{m_{1}}\\
w_{m_{0}}>w_{m_{1}}\\
w_{m_{0}}=w_{m_{1}}
\end{array}\end{cases}\hspace{-10mm}
\end{equation}
\end{prop}

Proposition 2 can be recursively used to find the value of AT. Fig.
3(a) shows the corresponding architecture for $\mathcal{L}=8$. It
consists of two radix-$\mathcal{L}/2$ sorters \cite{TVLSI_EPFL},
$\mathcal{L}-1$ MUXes, and $\log_{2}\mathcal{L}$ comparators. As
shown in Fig. 3(a), the $\mathcal{L}$ path metrics are evenly divided
into two groups and passed through the radix-$\mathcal{L}/2$ sorter.
As a result, the metrics in each group are sorted, i.e., $w_{0}^{3}\leq w_{1}^{3}\leq w_{2}^{3}\leq w_{3}^{3}$
and $w_{4}^{3}\leq w_{5}^{3}\leq w_{6}^{3}\leq w_{7}^{3}$. From Proposition
2, by comparing $w_{2}^{3}$ and $w_{6}^{3}$, the size of the median
candidate set is halved. In Fig. 3(a), the comparison result of $w_{2}^{3}$
and $w_{6}^{3}$ controls the 4 MUXes at stage 2 and they choose $\left\{ w_{0}^{2},w_{1}^{2},w_{2}^{2},w_{3}^{2}\right\} $
based on (10). Moreover, $w_{0}^{2}\leq w_{1}^{2}$ and $w_{2}^{2}\leq w_{3}^{2}$.
Hence, similar comparison and MUX architectures can be used for the
following stages. As a result, after $\log_{2}\mathcal{L}$ stages,
the median of the inputs, i.e., AT for the next depth, is obtained.

To find the value of RT, the architecture is simpler. If the maximum
path metric is adopted as RT as (8), the maximum of $w_{3}^{3}$ and
$w_{7}^{3}$ in Fig. 3(a) is RT. If the second maximum path metric
is taken for a tighter RT, it can be found by the architecture in
Fig. 3(b).
\begin{figure}
\begin{centering}
\includegraphics[scale=0.92]{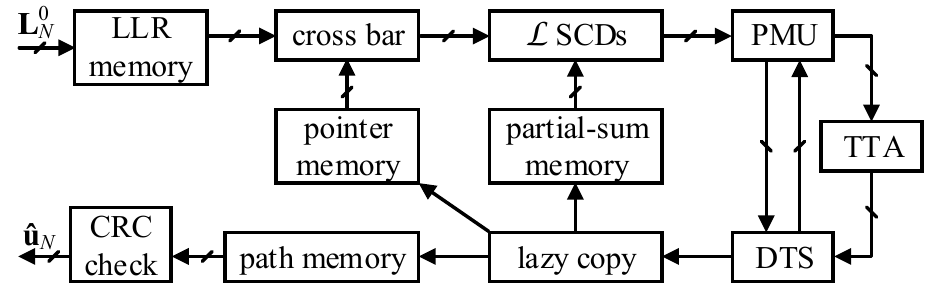}\vspace{-2mm}

\par\end{centering}

\caption{Block diagram of the low-latency list decoding architecture}
\vspace{-5mm}
\end{figure}

\subsection{List Decoding Architecture}

The top-level architecture of the proposed low-latency list decoding
is shown in Fig. 4. It contains $\mathcal{L}$ SCDs and each SCD is
implemented with a semi-parallel architecture of $M<N/2$ processing
elements (PEs) \cite{psn}. Based on $L_{i}^{n}$s output from the
SCD, the PMU generates $2\mathcal{L}$ $pm$s from $\mathcal{L}$
stored $pm$s with (4). Out of these $2\mathcal{L}$ $pm$s, $\mathcal{L}$
$pm$s are chosen by the DTS and they are stored in the register of
the PMU. Based on registered $pm$s, the TTA computes AT and RT used
in decoding $u_{i+1}$. After DTS, the memory contents related to
the SCD need to be copied as \cite{TCASII_EPFL}. As shown in Fig.
4, the lazy copy (LCP) block generates the control logic for them.
Finally, when the list decoding reaches the leaf node of the decoding
tree, the contents of the path memory are passed to the CRC check
block. The source word that satisfy the CRC check is the decoding
result $\mathbf{\hat{u}}_{N}$. 

\begin{figure}
\definecolor{lightgray}{gray}{0.6}\setlength\tabcolsep{2pt}\setlength{\extrarowheight}{2pt}\centering{}%
\footnotesize
\begin{tabular}{c"C{5mm}|C{5mm}|C{5mm}|C{5mm}|C{5mm}|C{5mm}|C{5mm}|C{5mm}|C{5mm}|C{5mm}|C{5mm}|C{5mm}} 
\hlinewd{1pt}  
CC & 0 & 1 & 2 & 3 & 4 & 5 & 6 & 7 & 8 & 9 & 10 & 11\tabularnewline 
\hlinewd{1pt} 
 & \cellcolor{lightgray}$f^1$ & \cellcolor{lightgray}$f^2$ & DTS & LCP & \cellcolor{lightgray}$g^2$ & DTS & LCP & \cellcolor{lightgray}$g^1$ & \cellcolor{lightgray}$f^2$ & DTS & LCP & \cellcolor{lightgray}$g^2$\tabularnewline 
\hline
 &  &  &  & TTA & TTA &  & TTA & TTA &  &  & TTA & TTA\tabularnewline
\hlinewd{1pt}  
\end{tabular}\vspace{-2mm}

\caption{Timing diagram of the low-latency list decoding architecture}
\vspace{-3mm}
\end{figure}
\begin{figure}[t]
\definecolor{lightgray}{gray}{0.6}\setlength\tabcolsep{2pt}\setlength{\extrarowheight}{2pt}\centering{}%
\footnotesize
\begin{tabular}{c"C{8mm}|C{8mm}|C{8mm}|C{8mm}|C{8mm}|C{8mm}|C{8mm}} 
\hlinewd{1pt}  
CC & 0 & 1 & 2 & 3 & 4 & 5 & 6 \tabularnewline 
\hlinewd{1pt} 
 & \cellcolor{lightgray}$f^1$ & PMU & \cellcolor{lightgray}$g^1$ & \cellcolor{lightgray}$f^2$ & DTS & LCP & \cellcolor{lightgray}$g^2$\tabularnewline 
\hline
 &  &  & TTA & TTA &  & TTA & TTA\tabularnewline
\hlinewd{1pt}  
\end{tabular}\vspace{-2mm}

\caption{List decoding timing diagram with frozen sibling}
\vspace{-5mm}
\end{figure}
Fig. 5 shows the timing diagram of the proposed low-latency list decoding
architecture, using the example in Fig. 1 for illustration. For simplicity,
assume there are already $\mathcal{L}$ decoding paths in the list
in the beginning. From Fig. 5, different from the conventional SCD,
two additional clock cycles are inserted after each leaf node SCD
operation of the scheduling tree. As depicted in Fig. 5, they are
used for the list pruning by DTS and the memory manipulation by LCP,
respectively. As a result, the latency of decoding one codeword in
terms of clock cycle number is given by
\begin{equation}
\tilde{T}_{d}=4N+\left(n-2-\log_{2}M\right)N/M.
\end{equation}
Finally, Fig. 5 also shows that the TTA is not on the critical path
of the list decoding. At least 2 clock cycles are available for the
TTA.

\subsection{Further Latency Reduction}

In this sub-section, frozen siblings are used to reduce the decoding
latency. They are defined as $\left[u_{2j},u_{2j+1}\right]$ with
$\left\{ 2j,2j+1\right\} \subset\mathcal{A}^{c}$. With $0\leq j<N/2$,
a frozen sibling corresponds to a leaf sibling in the scheduling tree.
For a general sibling, as shown in Fig. 5, $f^{n}$ and $g^{n}$ are
sequentially evaluated based on the LLR $\left[L_{2j}^{n-1},L_{2j+1}^{n-1}\right]$
of their parent in the scheduling tree. However, for a frozen sibling,
its path extension is fixed and given by $\left[\hat{u}_{2j},\hat{u}_{2j+1}\right]=\left[0,0\right]$.
Moreover, the PMU from $pm_{l}^{2j}$ to $pm_{l}^{2j+2}$ can be simplified
as
\begin{equation}
pm_{l}^{2j+2}=pm_{l}^{2j}+\Theta\left(L_{2j}^{n-1}\right)\left|L_{2j}^{n-1}\right|+\Theta\left(L_{2j+1}^{n-1}\right)\left|L_{2j+1}^{n-1}\right|.
\end{equation}
It can be proven that (12) is equivalent to (4) for the frozen sibling,
and 5 clock cycles can be saved by (12). For example, if $\left[u_{0},u_{1}\right]$
in Fig.1 is a frozen sibling, the timing diagram of the list decoding
is shown in Fig. 6. All the decoding operations related to the frozen
sibling shrinks to a PMU operation (12) in one clock cycle. Therefore,
the latency of the proposed list decoding is reduced to
\begin{equation}
T_{d}=4N+\left(n-2-\log_{2}M\right)N/M-5FS,
\end{equation}
where $FS$ is the number of frozen siblings in the given polar codes.

\section{EXPERIMENTAL RESULTS}

\label{sec:majhead}

\begin{figure}
\begin{centering}
\includegraphics[bb=20bp 0bp 720bp 540bp,scale=0.35]{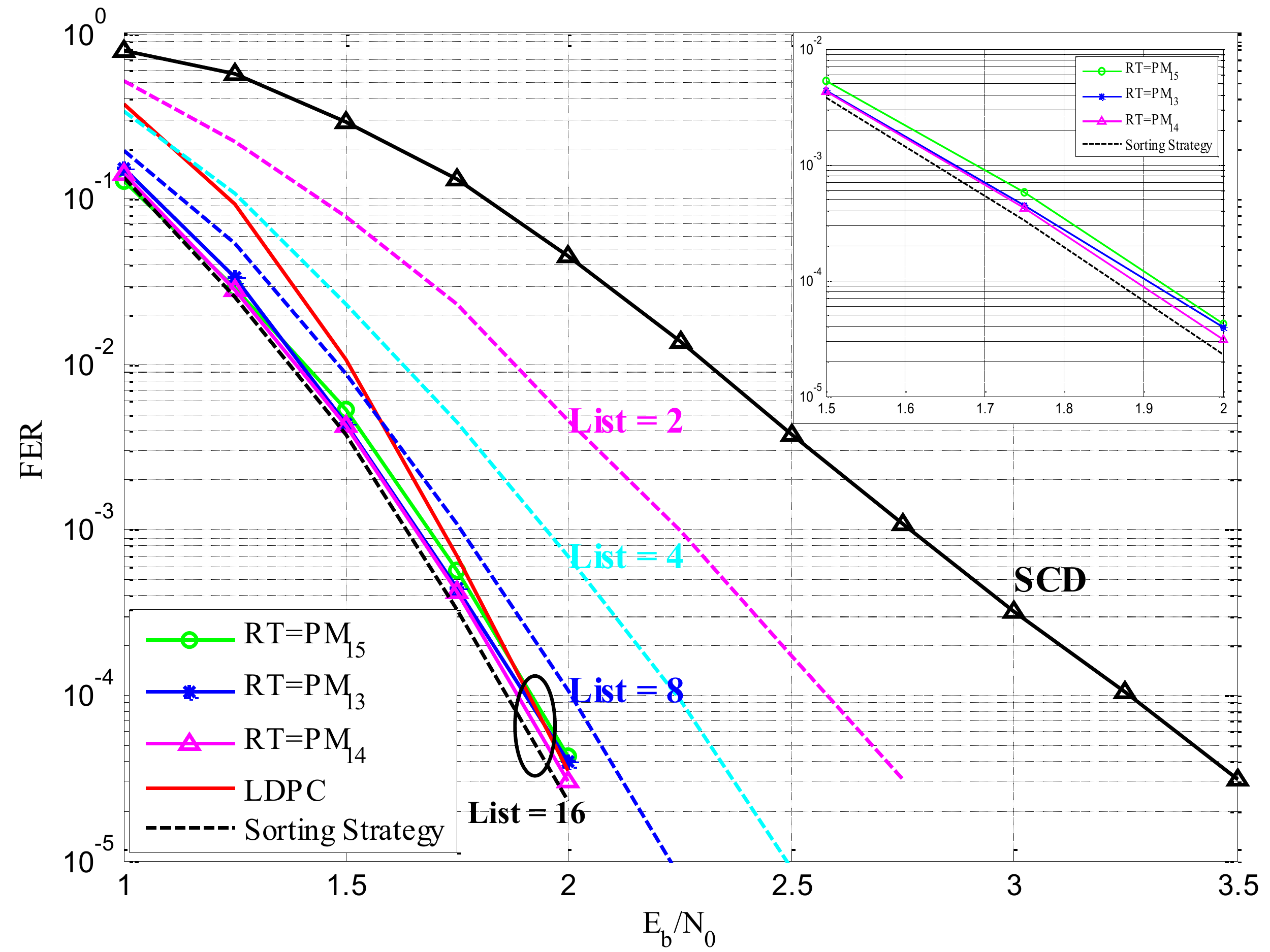}\vspace{-4mm}

\par\end{centering}

\caption{Performance comparison of different decoders}
\vspace{-3mm}
\end{figure}
\begin{table}[t]
\caption{\textsc{Synthesis Results and Comparison}}
\vspace{-3mm}

\definecolor{lightgray}{gray}{0.6}\setlength\tabcolsep{3pt}\setlength{\extrarowheight}{2pt}\centering{}%
\footnotesize
\begin{tabular}{c"c|c|c|c} 
\hlinewd{1pt}  
 & This Work & \cite{TCASII_EPFL} & \cite{ISCAS_Lehigh} & \cite{ICASSP_EPFL}\tabularnewline 
\hlinewd{1pt} 
Technology & \multicolumn{4}{c}{90 $nm$ CMOS}\tabularnewline 
\hline
LL/ LLR & LLR-based & LL-based & LL-based & LLR-based\tabularnewline 
\hline
$N$ & \multicolumn{4}{c}{1024}\tabularnewline
\hline
$\mathcal{L}$ & 16 & 4 & 8 & 4\tabularnewline
\hline
$M$ & \multicolumn{4}{c}{64}\tabularnewline
\hlinewd{1pt}
Area ($mm^2$) & 7.46 & 3.53 & 8.64 & 1.743\tabularnewline
\hline
Clock Freq. ($MHz$) & 641 & 314 & 625 & 412\tabularnewline
\hline
Throughput ($Mbps$) & 220 & 124 & 177 & 162\tabularnewline
\hlinewd{1pt}  
\end{tabular}\vspace{-5mm}
\end{table}
An $\left(N,R,r\right)=\left(2048,1/2,16\right)$ polar code is sent
over the BAWGN channel and decoded with different decoders. Their
frame error rate (FER) curves are shown in Fig. 7. All the list decodings
use (4) for PMU as in the hardware implementation. $\mathcal{L}=2,4,8$,
and 16 are simulated for conventional list decoding with sorting strategy
\cite{ICASSP_EPFL}. As a reference, the performances of the SCD and
$\left(N,R\right)=\left(2304,1/2\right)$ WiMAX LDPC code \cite{wimax}
are also presented. Here, 25 iterations are used for LDPC decoding.
It can be seen that the performance of polar codes is better than
that of the LDPC code, when $\mathcal{L}=16$ list decoding is used.
Finally, three DTSs are used for $\mathcal{L}=16$. Their $RT$s assume
$pm_{13}^{i}$, $pm_{14}^{i}$, and $pm_{15}^{i}$, respectively,
and $AT$ is fixed to $pm_{8}^{i}$. Fig. 7 indicates that $pm_{14}^{i}$
is the optimal value of $RT$ and the performance degradation of the
resulting low-latency list decoding is smaller than 0.02 dB.

The architecture shown in Fig. 4 is implemented for $\mathcal{L}=16$
to decode $\left(N,R\right)=\left(1024,1/2\right)$ polar codes. The
quantization scheme of \cite{ICASSP_EPFL} is used, i.e., 6 bits for
channel LLR $L_{i}^{0}$ and 8 bits for path metric. The design is
synthesized using a UMC 90 nm CMOS technology, and Table I summarizes
the synthesis results. Due to a large list size, the area of the LLR
memory in our implementation is large and equals to 4.5 $mm^{2}$.
For the target polar codes, $FS=231$ and decoding throughput can
be obtained from (13). From the table, the proposed architecture achieves
a decoding throughput of 220 Mbps, and it is even greater than that
of list size of 8 in \cite{ISCAS_Lehigh}. The results in Table I
demonstrate the effectiveness of the proposed low-latency list decoding
architecture with double thresholding.

\section{CONCLUSION}

\label{sec:print}

For a low-latency list decoding, a double thresholding strategy (DTS)
is proposed for fast list pruning. With a negligible performance degradation,
the DTS greatly reduces the pruning logic delay. Based on the DTS,
the low-latency list decoding architecture is proposed. Comparison
results demonstrate that the proposed architecture achieves a much
lower latency for a large list size.

\vfill{}
\pagebreak{}

\label{sec:refs}

 

\begin{thebibliography}{10}
\bibitem{Arikan} E.~Ar\i{}kan, ``Channel polarization: a method
for constructing capacity-achieving codes for symmetric binary-input
memoryless channels,''\emph{ IEEE Trans. Inform. Theory}, vol. 55,
no. 7, pp. 3051-3073, July 2009. 

\bibitem{SSC} A.~Alamdar-Yazdi and F.~R.~Kschischang, ``A simplified
successive-cancellation decoder for polar codes,'' \textit{IEEE Commun.}
\textit{Lett.}, vol. 15, no. 12, pp. 1378-1380, Dec. 2011.

\bibitem{ML_Gross} G.~Sarkis and W.~J.~Gross, ``Increasing the
throughput of polar decoders,'' \textit{IEEE Commun.} \textit{Lett.},
vol. 17, no. 4, pp. 725-728, Apr. 2013.

\bibitem{bypass_SSC} Z.~Huang, C.~Diao, J.~Dai, C.~Duanmu, X.~Wu,
and M.~Chen, ``An improvement of modified successive-cancellation
decoder for polar codes,'' \textit{IEEE Commun.} \textit{Lett.},
vol. 17, no. 12, pp. 2360-2363, Dec. 2013.

\bibitem{semi_parallel_Gross} C.~Leroux, A.~J.~Raymond, G.~Sarkis,
and W.~J.~Gross, ``A semi-parallel successive-cancellation decoder
for polar codes,'' \emph{IEEE Trans. Signal. Process.}, vol. 61,
no. 2, pp. 289-299, Jan. 2013.

\bibitem{look_ahead_parhi} C.~Zhang and K.~K.~Parhi, ``Low-Latency
sequential and overlapped architectures for successive cancellation
polar decoder,'' \emph{IEEE Trans. Signal. Process.}, vol. 61, no.
10, pp. 2429-2441, May 2013.

\bibitem{tcasi_parhi} B.~Yuan and K.~K.~Parhi, ``Low-Latency
successive-cancellation polar decoder architectures using 2-bit decoding,''
\emph{IEEE Trans. Circuits Syst. I, Reg. Papers}, vol. 61, no. 4,
pp. 1241-1254, Apr. 2014.

\bibitem{tcasii_parhi} C.~Zhang and K.~K.~Parhi, ``Latency analysis
and architecture design of simplified SC polar decoders,'' \emph{IEEE
Trans. Circuits Syst. II, Exp. Briefs}, vol. 61, no. 2, pp. 115-119,
Feb. 2014.

\bibitem{JSAC_Gross} G.~Sarkis, P.~Giard, A.~Vardy, C.~Thibeault,
and W.~J.~Gross, ``Fast polar decoders: algorithm and implementation,''
\textit{IEEE J. Select. Areas Commun.}, vol. 32, no. 5, pp. 946-957,
May 2014.

\bibitem{psn} Y.-Z.~Fan and C.-Y.~Tsui, ``An efficient partial-sum
network architecture for semi-parallel polar codes decoder implementation,''
\emph{IEEE Trans. Signal. Process.}, vol. 62, no. 12, pp. 3165-3179,
Jun. 2014.

\bibitem{SSCD_Gross} A.~J.~Raymond and W.~J.~Gross, ``A scalable
successive-cancellation decoder for polar codes,'' \emph{IEEE Trans.
Signal. Process.}, vol. 62, no. 20, pp. 5339-5347, Oct. 2014.

\bibitem{ASSCC} A.~Mishra, A.~J.~Raymond, L.~G.~Amaru, G.~Sarkis,
C.~Leroux, P.~Meinerzhagen, A.~Burg, and W.~J.~Gross, ``A successive
cancellation decoder ASIC for a 1024-bit polar code in 180 nm CMOS,''
in \textit{Proc. IEEE Asian Solid-State Circuits Conf. (A-SSCC)},
Nov. 2012, pp. 205-208.

\bibitem{list} I.~Tal and A.~Vardy, ``List decoding of polar codes,''
2012, arXiv:1206.0050v1 {[}Online{]}. Available: http://arxiv.org/abs/1206.0050v1

\bibitem{list_BUPT} K.~Chen, K.~Niu, and J.~R.~Lin, ``List successive
cancellation decoding of polar codes,'' \textit{Electron. Lett.},
vol. 48, no. 9, pp. 500-501, Apr. 2012.

\bibitem{stack_BUPT} K.~Niu and K.~Chen, ``Stack decoding of polar
codes,'' \textit{Electron. Lett.}, vol. 48, no. 12, pp. 695-697,
Jun. 2012.\vfill{}
\pagebreak{}

\bibitem{TC_BUPT} K.~Chen, K.~Niu, and J.~R.~Lin, ``Improved
successive cancellation decoding of polar codes,'' \textit{IEEE Trans.
Commun.}, vol. 61, no. 8, pp. 3100-3107, Aug. 2013.

\bibitem{SD_BUPT} K.~Niu, K.~Chen, and J.~R.~Lin, ``Low-Complexity
sphere decoding of polar codes based on optimum path metric,'' \textit{IEEE
Commun.} \textit{Lett.}, vol. 18, no. 2, pp. 332-335, Feb. 2014.

\bibitem{CRC_BUPT} K.~Niu and K.~Chen, ``CRC-Aided decoding of
polar codes,'' \textit{IEEE Commun.} \textit{Lett.}, vol. 16, no.
10, pp. 1668-1671, Oct. 2012.

\bibitem{CRC_Bin} B.~Li, H.~Shen, and D.~Tse, ``An adaptive successive
cancellation list decoder for polar codes with cyclic redundancy check,''
\textit{IEEE Commun.} \textit{Lett.}, vol. 16, no. 12, pp. 2044-2047,
Dec. 2012.

\bibitem{ICC_BUPT} K.~Niu, K.~Chen, and J.~R.~Lin, ``Beyond
Turbo codes: rate-compatible punctured polar codes,'' in \textit{Proc.
IEEE Int. Conf. Commun. (ICC)}, Jun. 2013, pp. 3423-3427.

\bibitem{TCASII_EPFL} A.~Balatsoukas-Stimming, A.~J.~Raymond,
W.~J.~Gross, and A.~Burg, ``Hardware architecture for list successive
cancellation decoding of polar codes,'' \emph{IEEE Trans. Circuits
Syst. II, Exp. Briefs}, vol. 61, no. 8, pp. 609-613, Aug. 2014.

\bibitem{ISCAS_Xiaohu} C.~Zhang, X.~You, and J.~Sha, ``Hardware
architecture for list successive cancellation polar decoder,'' in
\textit{Proc. IEEE Int. Symp. Circuits Syst. (ISCAS)}, Jun. 2014,
pp. 209-212.

\bibitem{ISCAS_Lehigh} J.~Lin and Z.~Yan, ``Efficient list decoder
architecture for polar codes,'' in \textit{Proc. IEEE Int. Symp.
Circuits Syst. (ISCAS)}, Jun. 2014, pp. 1022-1025.

\bibitem{ICASSP_EPFL} A.~Balatsoukas-Stimming, M.~B.~Parizi, and
A.~Burg, ``LLR-Based successive cancellation list decoding of polar
codes,'' in \textit{Proc. IEEE Int. Conf. Acoust., Speech, Signal
Process. (ICASSP)}, May 2014, pp. 3903-3907.

\bibitem{Asilomar_Parhi} B.~Yuan and K.~K.~Parhi, \textquotedbl{}Successive
cancellation list polar decoder using log-likelihood ratios,\textquotedbl{}
presented in \textit{Asilomar Conf.}, 2014, arXiv:1411.7282 {[}Online{]}.
Available: http://arxiv.org/abs/1411.7282

\bibitem{arxiv_EPFL} J.~Lin, C.~Xiong, and Z.~Yan, ``A reduced
latency list decoding algorithm for polar codes,'' 2014, arXiv:1405.4819v1
{[}Online{]}. Available: http://arxiv.org/abs/1405.4819v1

\bibitem{arxiv_Gross} G.~Sarkis, P.~Giard, A.~Vardy, C.~Thibeault,
and W.~J.~Gross, ``Increasing the speed of polar list decoders,''
2014, arXiv:1407.2921v1 {[}Online{]}. Available: http://arxiv.org/abs/1407.2921v1

\bibitem{TVLSI_Parhi} B.~Yuan and K.~K.~Parhi, \textquotedblleft{}Low-latency
successive-cancellation list decoders for polar codes with multibit
decision,\textquotedblright{} \emph{IEEE Trans. Very Large Scale Integr.
Syst.}, to appear.

\bibitem{TVLSI_EPFL} L.~Amaru, M.~Martina, and G.~Masera, ``High
speed architectures for finding the first two maximum/minimum values''
\emph{IEEE Trans. Very Large Scale Integr. Syst.}, vol. 20, no. 12,
pp. 2342-2346, Dec. 2012.

\bibitem{wimax} \textit{Air Interface for Fixed and Mobile Broadband
Wireless Access Systems}, IEEE 802.16e, Oct. 2005 {[}Online{]}. Available:
http://www.ieee802.org/16/tge

\end{thebibliography}
\end{document}